\documentclass[twocolumn]{aastex631}

\usepackage{amsmath}
\usepackage{CJKulem}

\begin{document}
\shortauthors{Zhong \& Yu}

\title{In situ formation of super-Earth/sub-Neptune driven by the planetary rotation}

\correspondingauthor{Cong Yu}
\email{yucong@mail.sysu.edu.cn}
\author{Wei Zhong}
\affiliation{School of Physics and Astronomy, Sun Yat-Sen University, Zhuhai, China, 519082}
\affiliation{CSST Science Center for the Guangdong-Hong Kong-Macau Greater Bay Area, Zhuhai, China, 519082}

\author{Cong Yu}
\affiliation{School of Physics and Astronomy, Sun Yat-Sen University, Zhuhai, China, 519082}
\affiliation{CSST Science Center for the Guangdong-Hong Kong-Macau Greater Bay Area, Zhuhai, China, 519082}


\begin{abstract}

Kepler's observation shows that many of the detected planets are super-Earths. They are inside a range of critical masses overlapping the core masses  (2-20 $M_{\bigoplus}$),  which would trigger the runaway accretion and develop the gas giants. Thus, super-Earths/sub-Neptunes can be formed by restraining runaway growth of gaseous envelopes.   We assess the effect of planetary rotation in delaying the mass growth. The centrifugal force, induced by spin, will offset a part of the gravitational force and deform the planet. Tracking the change in structure, we find that the temperature at the radiative-convective boundary (RCB) is approximate to the boundary temperature. Since rotation reduces the radiation energy densities in the convective and radiative layers, RCB will penetrate deeper.  The cooling luminosity would decrease. Under this condition, the evolutionary timescale can exceed the disk lifetime (10 Myr), and a super-Earth/sub-Neptune could be formed after undergoing additional mass loss processes. In the dusty atmosphere, even a lower angular velocity can also promote a super-Earth/sub-Neptune forming. Therefore, we conclude that rotation can slow down the planet's cooling and then  promote a super-Earth/sub-Neptune forming.

\end{abstract}


\section{Introduction} \label{sec:intro}

In the formation of planets, rapid rotation plays a crucial role, and it is not a small perturbation on spherically symmetric structures \citep{2018Natur.555..227G, 2018Natur.555..220I,2019sf2a.conf..453H}, since centrifugal force causes deformation to occur \citep{2000ssd..book.....M,2016ApJ...826..127K}. Besides, the shape and interior structure are uncertain in rotating planets, making the determination of the interior structure and evolution of planets a challenge.

Gravity measurements taken by the Juno \citep{2010GeoRL..37.1204K} and Cassini  \citep{2019Sci...364.2965I} spacecraft provided precise data of even gravitational harmonics to constrain the interior structure of Jupiter and Saturn \citep{2020A&A...639A..10N}, respectively, under the effect of rotational distortion. However, the predicted gravitational harmonics using a typical rigid-body model \citep{1978ppi..book.....Z,Guillot1995CEPAM,2005AREPS..33..493G,2020A&A...639A..10N} were lower than data from these measurements. Therefore, other mechanisms may also influence gravitational harmonics. The contribution of deep zonal flow may be a possible explanation for the higher gravitational harmonic data supplied by the Juno space probe \citep{dali2013A,2015MNRAS.450L..11K}. Wind velocity \citep{2019GeoRL..46..616G,2019Sci...364.2965I} can influence Saturn's shape and interior structure, and when the wind extends to critical depth ($\sim 8800\,km$), the calculated results match the actual measurements \citep{2019GeoRL..46..616G,2019Sci...364.2965I}.
The metallic dynamo region in the rotational Saturn can also affect the gravitational field by constraining the strong zonal circulation \citep{2019MNRAS.488.5633K}. 
\cite{2017GeoRL..44.4649W} assert that the dilute core helps reconcile the calculated gravitational harmonics with high metallic element content deep below the surface of Saturn. 

Spin evolution determines how planets are formed. If the early angular moment is conserved, planets will rotate at near break-up velocity and not contract \citep{2020MNRAS.491L..34G}. However, magnetic braking \citep{2018AJ....155..178B,2020MNRAS.491L..34G} or tidal force \citep{1968ARA&A...6..287G} reduces the spin velocity, consequently ending this condition. This is consistent with the measurements of rotation lines in eight planetary-mass objects by high-resolution near-infrared spectra from NIRSPEC/Keck  \citep{2020ApJ...905...37B}. The spin velocity is below the break-up velocity, and the constraint of spin velocity corresponds with the planetary radius \citep{2020ApJ...905...37B}. 

About a fifth of Sun-like stars host super-Earths located within 0.05 - 0.3 AU with radii in the range of 1 to 4 times the radius of the Earth  \citep{2010Sci...330..653H,2013ApJS..204...24B,2013ApJ...770...69P}. The masses of these super-Earths are between 2 and 20 times the mass of the Earth and have been measured by transit timing variations \citep{2013ApJ...772...74W} and radial velocity measurements \citep{2014ApJ...783L...6W}. This mass range overlaps the range of core masses that might trigger gas runaway, meaning these super-Earths may grow to be gaseous giant planets. How these super-Earths avoid becoming gas giants is an intriguing issue.

Many factors can influence the formation of super-Earths, which tend to form in gas-poor or dusty atmospheres \citep{2014ApJ...797...95L,2015ApJ...811...41L}. For instance, the flow of disk gas penetrates the interior envelope with high entropy (i.e., entropy advection) and slows down the cooling of the atmosphere \citep{2020Ali}.  \cite{2020ApJ...896..135C} have also verified that the runaway growth may be self-impeded by an increase in grain opacity during the pebble-isolation phase and then develop super-Earths. \cite{2017ApJ...850..198Y} also demonstrated that tidally-forced turbulence promotes the formation of super-Earths with the inward movement of RCBs. The formation of super-Earths is related to the loss of mass. 
Almost $90\%$ of the gas envelope can be lost during the giant impact process \citep{2016ApJ...817L..13I,2015MNRAS.446.1685L,2019The}. As a result of these mechanisms, super-Earths can form.

In this paper, we investigated the influence of rotation on the long-term evolution of planets. 
Planetary spin alters the gravity potential with centrifugal force and deformation. Thus, the hydrostatic equilibrium and thermal gradient change during the slow gas accretion phase. Spin may have a crucial effect on the accretion history of a planet, as it would reduce the cooling luminosity and enhance the KH time. Based on our calculations, we assert that planetary spin can promote the formation of super-Earths/sub-Neptunes.

 Rotation would delay the epoch of runaway accretion in the early formation stage. When the disk gas is depleted, a mass fraction of a rotating planet is greater than 50 \%, much higher than the typical super-Earth, a sub-Saturn will be formed. It is difficult for rotation alone to reduce the mass fraction to be consistent with observation. Here we stress that the planet during the accretion stage may not be the final form of the planet revealed by observations. The planet may experience significant mass losses during the post-formation stage, due to gravitational tides, photoevaporation, giant impact, etc. The close-in planet undergoing the post-formation evolution (after the dispersal of the gas disc) experiences significant mass loss by photoevaporation  \citep{2016ApJ...816...34O,2017ApJ...847...29O,2019ApJ...872...99G}. Thus, the observed atmospheric masses are not the same as those accreted in the presence of the primordial gas disc \citep{2012ApJ...753...66I,2016ApJ...816...34O,2021MNRAS.503.5658M}. The giant impact \citep{2016ApJ...817L..13I,2015MNRAS.446.1685L,2019The}, core-powered effect  \citep{2019MNRAS.487...24G,2020MNRAS.493..792G,2021MNRAS.503.5658M}, and tidal effect can strip off the primal envelope. When these processes are considered, the mass of the planet may lose a large fraction of the its masses. As a result, we may obtain a super-Earth/sub-Neptune. In contrast, when the planet experiences the gas runaway accretion and becomes a gas giant, it is difficult for the mass losses to reduce the planet  mass to the range of the super-Earth/sub-Neptune. However, the planet that avoids the gas runaway phase may becomes a super-Earth or a sub-Neptune according to the mass loss it experiences. In other words, the stopping mechanism of the gas runaway plays an important role in determining the fate of the planet. In this paper, we use the term super-Earth and sub-Neptune interchangeably.

We start in Section \ref{sec:snapshot} with the description of the structure of a rotating planet.  In section \ref{results}, we show the effects of rotation in the radial profiles and the evolutionary processes. We summarize and discuss the formation of super-Earths/sub-Neptunes in Section \ref{sec:conclusion}.

\section{The structure of a rapidly rotating planet}\label{sec:snapshot}

Theoretically, the centrifugal force driven by rapid rotation will reduce the effective gravity and distort the planetary shape. Thus, the spherical symmetry structure would not be compatible.  \cite{1997A&A...321..465M,2009PhT....62i..52M} provided a shellular rotation model with constant angular velocity to solve the deformation and centrifugal force problem. \cite{2002A&A...394..965Z} provided a useful method to simplify the shellular rotation \citep{1997A&A...321..465M} and found their evolutionary results are similar.

Following \cite{2002A&A...394..965Z}, we find that an isometric equivalent sphere can show the features of the equipotential surface of the planet. The equivalent sphere is on the isobar (i.e., the constant pressure surface). Assuming the planet rotates at a fixed velocity, we can get a workable point to investigate the planetary structure. The planet's interior is governed by the equations of mass conservation, hydrostatic equilibrium, and thermal gradient \citep{2002A&A...394..965Z}:
\begin{equation}
  \frac{{\rm d} M}{{\rm d} r}=4 \pi \rho r^{2} \langle f_{\rm d} \rangle ,
  \label{mass_structure}
\end{equation}
\begin{equation}
  \frac{{\rm d} P}{{\rm d} r}=-\frac{G M}{r^{2}} \rho \langle f_{\rm d} \rangle  f_{\rm p },
  \label{hydrostatic_equilibrium}
\end{equation}
\begin{equation}
  \frac{{\rm d} T}{{\rm d} r}=\nabla \frac{T}{P} \frac{{\rm d} P}{{\rm d} r}.
  \label{thermal_gradient}
\end{equation}
where, mass, pressure, density, and temperature correspond to $M$, $P$, $\rho$, and $T$, respectively. The symbols $r$ is radius, $\nabla$ the thermal gradient and $G$ the gravitational constant.  Besides, the ideal gas defines $P= k_{\rm B} \rho T /\mu m_{\rm u}$ with the molecular weight $\mu$, Boltzmann constant $k_{\rm B}$, and the atomic mass unit $m_{\rm u}$. 

In the works of \cite{1997A&A...321..465M,2009PhT....62i..52M}, the average density $\overline{\rho}$ represents the density over an isobar surface, since the value of density is not a constant. \cite{2002A&A...394..965Z} employ the mean value $\left \langle f_{\rm d}\right \rangle$ to change the density. We can see $\left \langle f_{\rm d}\right \rangle$ as follows
\begin{equation}
\left \langle f_{\rm d}\right \rangle = \frac{\left ( 1-r^2 {\rm sin}^2 \theta \omega \alpha \right )\left \langle g_{\rm eff}^{-1}\right \rangle}{\left \langle g_{\rm eff}^{-1}\right \rangle-\left \langle g_{\rm eff}^{-1} r^2 {\rm sin}^2 \theta  \right \rangle  \omega \alpha},
\label{f_d}
\end{equation}
where, $\omega$ is the angular velocity, $\theta$ denotes the colatitude, and $g_{\rm eff}$ corresponds to the effective gravitational acceleration. As the angular velocity is fixed in the rigid model and the parameters are the constant in the equivalent sphere, $\alpha = {\rm d}\omega/{\rm d}\Psi =1$ ($\Psi$ is potential). Thus, $\left \langle f_{\rm d}\right \rangle$  can be set to 1.

The parameter of $f_{\rm p}$ \citep{2002A&A...394..965Z}, determined by the effective gravity, gravity, and centrifugal accelerations (i.e.,  $g_{\rm eff}$, $g_{\rm r}$, and $a_{\rm n}$ ), can be expressed by
\begin{equation}
  f_{\rm p} = \frac{1}{2} \frac{g_{\rm r}^{2}+g_{\rm eff}^{2}-a_{\rm n}^{2}}{g_{\rm r}^{2}},
\label{f_p}
\end{equation}
where, the effective gravitational acceleration satisfies  \citep{2002A&A...394..965Z,2009PhT....62i..52M}
\begin{equation}
 g_{\rm eff} =  \left[\left(-g_{\rm r}+a_{\rm n} \sin \theta\right)^{2}+ \left(a_{\rm n} \cos \theta\right)^2 \right] ^{1/2}.
\label{g_eff}
\end{equation}
The centrifugal acceleration corresponds to $a_{\rm n}= \omega^{2} r \sin \theta$, while the gravitational acceleration is $g_{\rm r} = GM/r^2$.

The thermal gradient is determined by the radiative gradient ($\nabla_{\rm rad}$) and adiabatic gradient ($\nabla_{\rm ad }$), i.e., $\nabla=\min \left(\nabla_{\rm ad},\nabla_{\rm rad}\right)$.  The radiative-convective boundary can be found at $\nabla_{\rm ad}=\nabla_{\rm rad}$. A smaller adiabatic gradient ($\nabla_{\rm ad } = 0.17$) is adopted, the RCB would shift outward. As a result, the evolutionary timescale can also be decreased, which is compatible with the comparison between the left and right top panels in Figure 6 from  \cite{2020ApJ...896..135C}. Therefore, we can predict that the RCB of a rotating planet will move outwards driven by a lower adiabatic gradient, which will suppress the effect of rotation.  As we mentioned in the above response, our treatment of rotation is highly simplified, we need to perform more rigorous treatment of the rotation to understand further about the effects on the early accretion history of planet. In addition, a very efficient mechanism to enhance the accretion timescale called entropy advection is recently proposed. How does the nonlinear interaction between the rotation and entropy advection \citep{2020Ali} influence the early evolution of planet accretion is also worth a further investigation. The radiative gradient 
\begin{equation}
    \nabla_{\rm rad}=\frac{3 \kappa L P}{64 \pi \sigma G M_{r}  T^{4}} f_{\rm R},
    \label{rad_gradient}
\end{equation}
where $\sigma$ is the Stefan-Boltzman constant and $L$ is the luminosity. The coefficient $f_{R}$ satisfies
\begin{equation}
f_{\rm R} = \frac{g_{\rm r}}{g_{\rm eff}} \frac{4 \pi r^{2}}{S_{\rm P}} = \frac{g_{\rm r}}{g_{\rm eff}} \frac{3 r^{2}}{2a_{\rm e}^2+b^2},
\label{f_r}
\end{equation}
where $S_{\rm P}$ is the surface area of the isobar.

Generally, $f_{\rm R}=f_{\rm p}=1$ when the planet evolves without rotation.  To derive the values of $f_{\rm R}$ and $f_{\rm p}$ in the rotational state, we need to evaluate the colatitude at a fixed angular velocity. The colatitude $\theta$ can be determined by the polar equation $b$ and radius $r$ on the isobar, and it can be shown as follows \citep{2009PhT....62i..52M}
\begin{equation}
\frac{G M}{r}+\frac{1}{2} \omega^{2} r^{2} \sin ^{2} \theta=\frac{G M}{b},
\label{theta}
\end{equation}
where, $b = r(f-\eta)^{\frac{2}{3}} $   with $\eta  = {\omega^{2} r^{3}}/{2G M}$ and the non-rotating parameter $f$. The equatorial radius $a_{\rm e} = b/(f-\eta)$. As the angular velocity  vanishes, the parameter $f=1$.  Note that the critical angular velocity $\omega_{\rm crit}^2 = GM/a_{\rm e}^3$ controls the maximal rotational velocity.

The opacity of gas follows \citep{2015ApJ...811...41L,2017ApJ...850..198Y}
\begin{equation}
\kappa = \kappa_{\rm 0}\left(\frac{P}{P_{\rm 0}}\right)^{a} \left(\frac{T}{T_{\rm 0}}\right)^{b},
\label{kappa_gas}
\end{equation}
Here, we choose $\kappa_{\rm 0 }= 0.01 \,cm^{2}\,g^{-1}$. To ensure the existence of inner convective layer and outer radiative layer, we set $a=1$ and $b=1$ \citep{2006ApJ...648..666R}. $P_{\rm 0}$ and $T_{\rm 0}$ are determined by the outer pressure and temperature, respectively.
When the dusty grains are included, the opacity can split into two templates \citep{2014Ormel}:
\begin{equation}
\kappa = \kappa_{\rm gas}+\kappa_{\rm gr}=\kappa_{\rm gas} + \kappa_{\rm geom}Q_{\rm e},
\label{kappa_tot}
\end{equation}
where the gas opacity $\kappa_{\rm gas}$ follows Equation (\ref{kappa_gas}), and the grain opacity corresponds to $\kappa_{\rm gr}$. Besides, the geometrical opacity  is defined by the grain abundance $Z$ and Monomer grain radius $s_{\rm 0 } = 1 \,um$ \citep{2014Ormel}, i.e., $\kappa_{\rm geom}=3Z/4 \rho_{0}s_{\rm 0}$, with the grain density $\rho_{\rm 0} = 3 \,g\, cm^{-3}$. Remarkably, the efficiency factor 
\begin{equation}
Q_{\rm e} = \min\left(0.3x,2\right),\, x=\frac{2\pi s_{\rm 0}}{\lambda_{\rm max}\left(T\right)},
\label{Q}
\end{equation}
where, $\lambda_{\rm max} \left(T\right)$ is the peak wavelength from Wien's displacement law.

We solve Equations (\ref{mass_structure}-\ref{thermal_gradient}) from the outer to the interior to get the planet's structure. The outer boundary is regulated by the minimum value between the Bondi and Hill radius \citep{1986Bodenheimer}, i.e., $R_{\rm out}=\min\left(R_{\rm B},R_{\rm H}\right)$. Especially, the Hill radius is $R_{\rm H}=r_{\rm au} \left(M_{\rm p}/3M_{\star }\right)^{1/3}$ and the Bondi radius is $R_{\rm B}=GM_{\rm p}/c_{\rm s}^2$, where $r_{\rm au}$ is the orbital radius, $M_{\rm p}$ is the total mass, $M_{\star}$ is the mass of sun, and $c_{\rm s}^2=\gamma k_{\rm B}T_{\rm d}/\mu$ ($\gamma = 1.4$) is the  sound speed. Additionally, the minimum-mass extra-solar nebula (MMEN,\citealt{2013Chiang}) model has the outer density 
\begin{equation}
 \rho_{\rm d}= 7.6 \times 10^{-9} \mathrm{~g} \mathrm{~cm}^{-3} r_{\rm au}^{-2.9},
 \label{out_rho}
\end{equation}
and the temperature
\begin{equation}
  T_{\rm d}=373 \mathrm{~K} \,\, r_{\rm au}^{-3/7}.
  \label{out_t}
\end{equation}

Time steps can be estimated by the difference between any two adjacent steady states as \citep{2013Piso}
\begin{equation}
\Delta t = \frac{-\Delta E + \left \langle e \right \rangle \Delta M - P \Delta V_{\left \langle M \right \rangle}}{\left \langle L\right \rangle},
\label{time_step}
\end{equation}
where $\left \langle {\rm x}\right \rangle$ represents the mean value of quantity ${\rm x}$ in two adjacent states, while $\Delta $ is a difference between two states. The total energy equals to the sum of gravitational and internal energy, i.e., $E=E_{\rm G}+U=-\int {G M}/{r} \,{\rm d} M +\int u \,{\rm d} M$  with the specific internal energy
$u=C_{\rm v} T=\left(\nabla_{\rm ad}^{-1}-1\right) T \, {k_{\rm B}}/{m_{\mu}}$. The volume difference $\Delta V_{\langle M\rangle}$ is performed at the average mass between two states \citep{2013Piso}.

\begin{figure*}[!htp]
\gridline{\fig{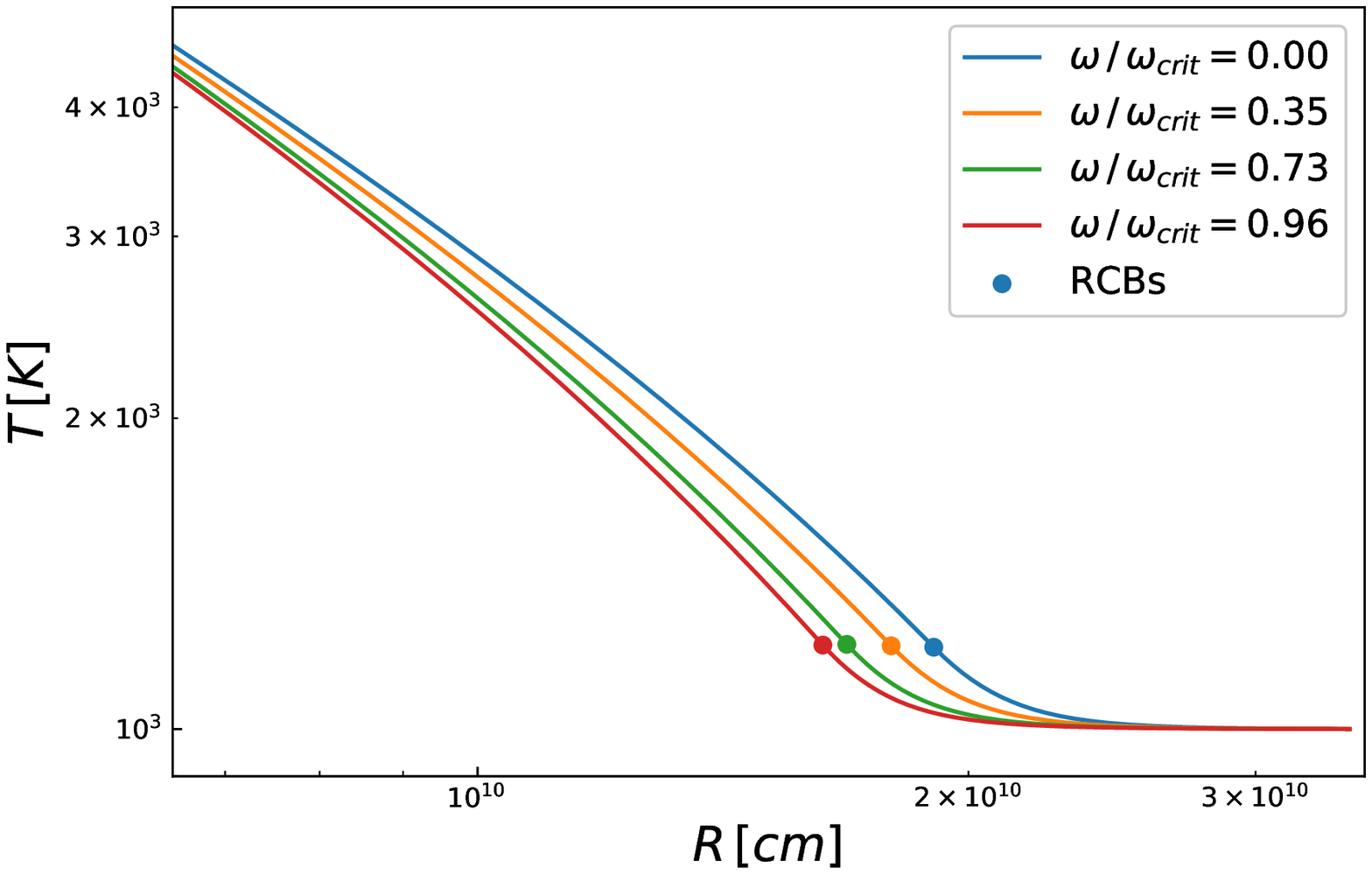}{0.45\textwidth}{}
          \fig{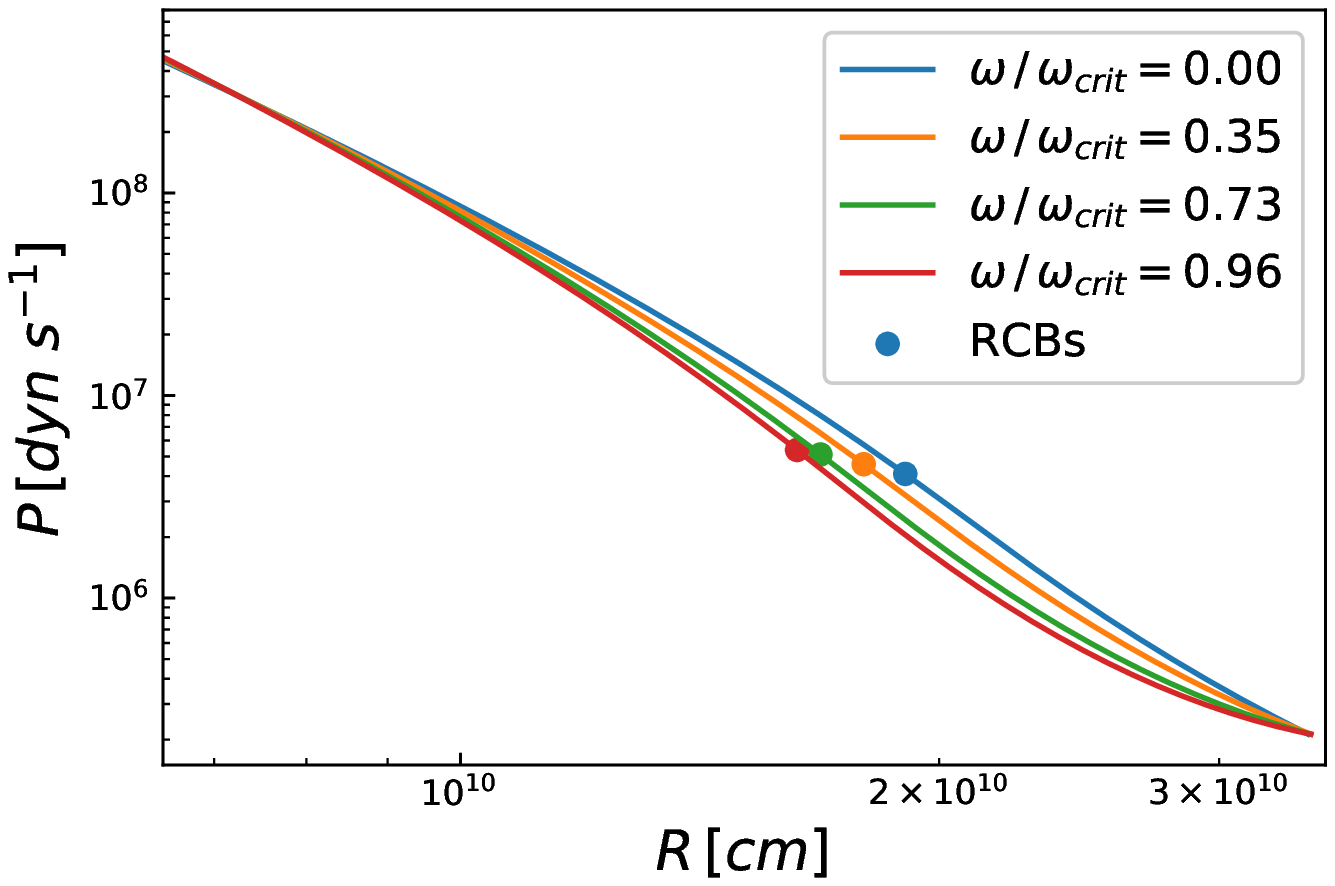}{0.45\textwidth}{}}\vspace{-1.0cm}
\gridline{\fig{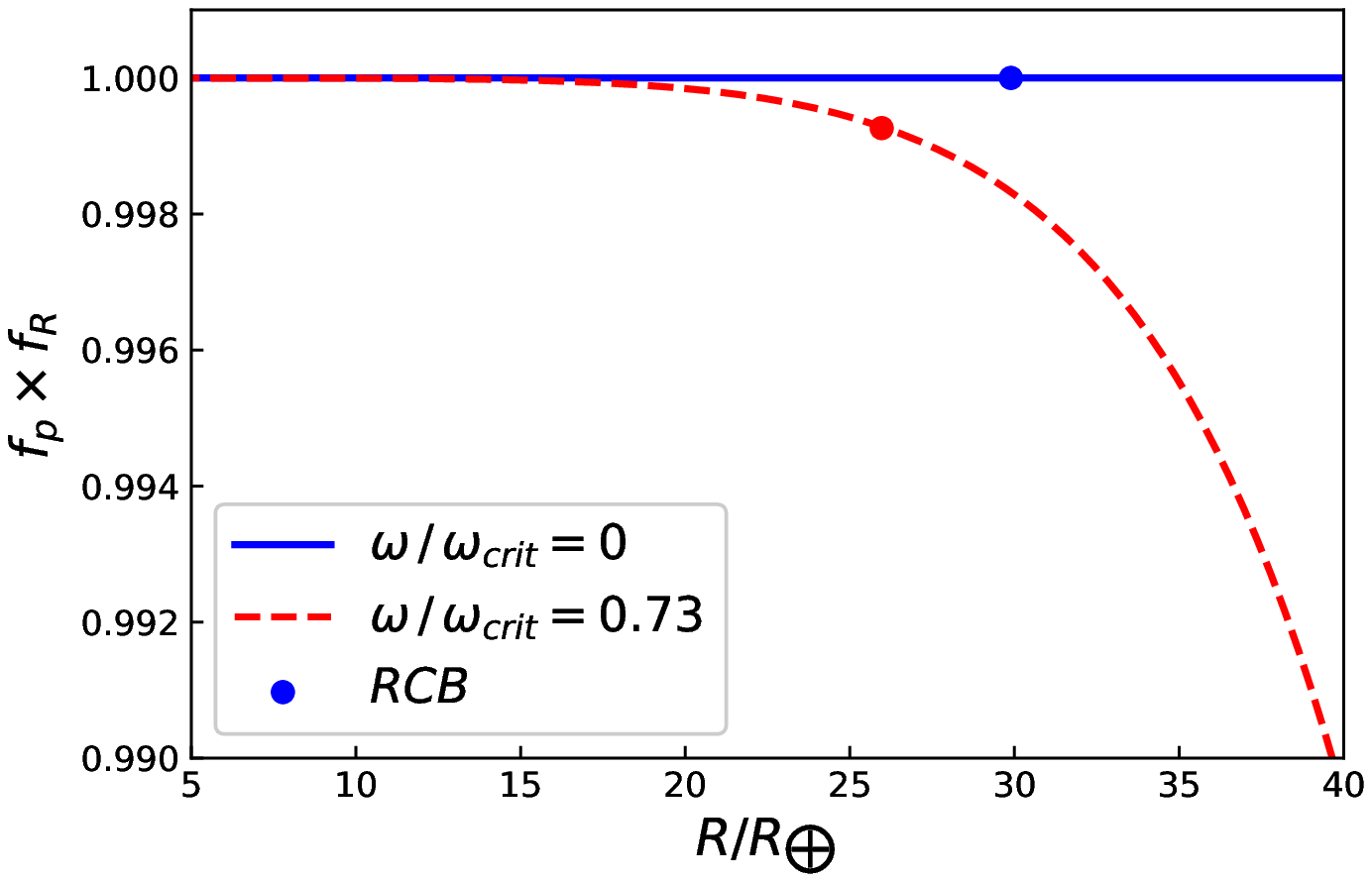}{0.45\textwidth}{}
          \fig{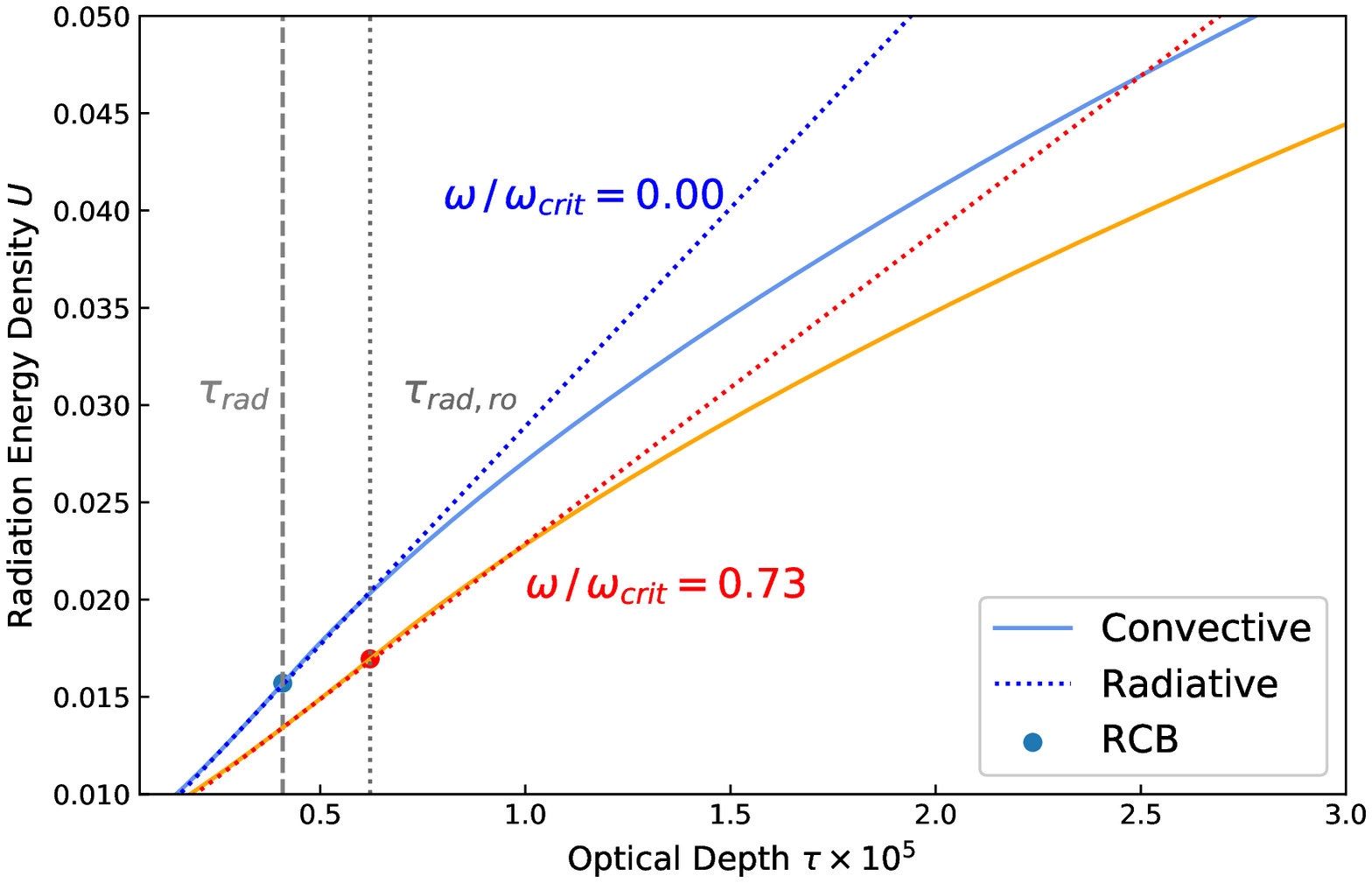}{0.45\textwidth}{} }\vspace{-0.8cm}
\caption{The upper panels show the radial profiles of temperatures and pressures at the same gas-to-core mass ratio GCR ($M_{\rm atm}/M_{\rm c}$ = 0.5), respectively. Different colors correspond to the different angular velocities, which decrease from left to right. The lower panels are for the radial profiles of the rotational coefficient ($f_{\rm p}f_{\rm R}$) and radiation energy densities, respectively. Blue and red lines indicate the radial curves without and with rotation, respectively. The dots denote the locations of RCBs. In the lower right, the solid and dotted lines are for radial curves of the radiation energy densities in convective and radiative layers, respectively.  Gray dashed and dotted lines represent the optical depth of the radiative-convective transition point without and with rotation (i.e., $\tau_{\rm rad}$ and $\tau_{\rm rad,ro}$), respectively.}
\label{rotational_structure}
\end{figure*}

\section{Results}\label{results}
In this paper, we focus on whether the  super-Earths/sub-Neptunes form by rotation. The radial profiles  of a rotating planet are shown in \ref{three_layer}. The formation of a   super-Earth/sub-Neptune  is shown in Section \ref{Super-Earth}.

\subsection{The Radial Profiles of A Rotating Planet}\label{three_layer}

For convenience, we set the fiducial model without rotation and the rotational model in the same figure. Figure \ref{rotational_structure} shows the radial profiles at the same gas-to-core mass ratio (i.e., $GCR=M_{\rm atm}/M_{\rm core}$) with the different angular velocities, i.e., $\omega/\omega_{\rm crit}=$ 0.00, 0.35, 0.73, and 0.96, respectively. The core mass $M_{\rm core}$ and radius $R_{\rm core}$ are set to $8\,M_{\bigoplus}$ and $2.0\,R_{\bigoplus}$ within $0.1$ AU, respectively.

Rotation changes the parameters at RCBs, especially for the temperature and pressure (i.e., $P_{\rm RCB}$, $T_{\rm RCB}$), and then works on the planetary evolution. We relate $P_{\rm RCB}$ and $T_{\rm RCB}$ to the surface temperature $T_{\rm 0}$ and pressure $P_{\rm 0}$, respectively. The outward flux \citep{2006ApJ...650..394A} in the radiative zone carried by radiative diffusion is 
\begin{equation}
\begin{aligned}
F = &-\frac{16\sigma T^3}{3\kappa \rho} \frac{{\rm d}T}{{\rm d}r} =
\frac{16\sigma T^3 g_{\rm r} f_{\rm p}}{3\kappa} \frac{{\rm d}T}{{\rm d}P}\\
=& \nabla_{\infty}\frac{16\sigma g_{\rm r} f_{\rm P}}{3\kappa_{\rm 1}}\frac{{\rm d}T^{4-b}}{{\rm d}P^{a+1}},
\end{aligned}
\label{out_flux}
\end{equation}
where, we assume the constant gravity $g_{\rm r} = GM/r^2$, pressure $P=k_{\rm B} \rho T/\mu m_{\rm u}$, the opacity $\kappa=\kappa_{\rm 0} \left(P/P_{\rm 0}\right)^{a} \left(T/T_{\rm 0}\right)^{b} = \kappa_{\rm 1} P^{a} T^{b}$. The temperature gradient for a radiative zero solution defines $\nabla_{\infty} = \left(a+1\right)/\left(4-b\right)$. Note that the coefficient $f_{\rm p} = 1-a_{\rm n}\sin\theta/g_{\rm r}<1$ according to Equation (\ref{f_p}). Integrating Equation (\ref{out_flux}), we find
\begin{equation}
T^{4-b} = {\rm constant} + f_{\rm p} \nabla_{\infty}^{-1} \left(\frac{3\kappa_{\rm 1}F}{16\sigma g_{\rm r}} \right)P^{a+1}.
\label{eq:(24)}
\end{equation}
Since $T \simeq T_{\rm 0}$ at small pressure, temperature profile will switch into
\begin{equation}
\begin{aligned}
T = & T_{\rm 0}\left[1+\left(\frac{3\kappa_{\rm 1}F}{16\sigma g_{\rm r}\nabla_{\infty} T_{\rm 0}^{4-b}}\right) P^{a+1} f_{\rm p}
\right]^{1/\left(4-b\right)}\\
= & T_{\rm 0}\left[1+\left(P/P_{\rm 0}\right)^{a+1} f_{\rm p}
\right]^{1/\left(4-b\right)},
\end{aligned}
\label{temperature_an}
\end{equation}
where the surface pressure $P_{\rm 0} = \left(16 \sigma g_{\rm r} T_{\rm 0}^{4-b} \nabla_{\infty}/{3\kappa_{\rm 1}F}\right)^{1/\left(a+1\right)}$. Thereby, the logarithmic temperature gradient is given by
\begin{equation}
\nabla = \nabla_{\infty} f_{\rm p} \frac{\left(P/P_{\rm 0}\right)^{a+1}}{1+\left(P/P_{\rm 0}\right)^{a+1}}.
\label{gradient_an}
\end{equation}

To solve for the transition of convective and radiative layers, we set $\nabla=\nabla_{\rm ad}$. Thus, the pressure and temperature at RCB are written as
\begin{equation}
P_{\rm RCB} = P_{\rm 0}\left[\frac{\nabla_{\rm ad}}{\left(\nabla_{\infty}-\nabla_{\rm ad}\right)f_{\rm p}}\right]^{1/\left(a+1\right)},
\label{p_rcb}    
\end{equation}
\begin{equation}
T_{\rm RCB} = T_{\rm 0}\left(1+\frac{\nabla_{ \infty}}{\nabla_{\infty}-\nabla_{\rm ad}}\right)^{1/\left(4-b\right)}.
\label{T_rcb}  
\end{equation}
When $f_{\rm p}=f_{\rm R}=1$, the planet grows without the effect of rotation. Following Equation (\ref{p_rcb}), we hold that the change of $f_{\rm P}$ determine the values of $P_{\rm RCB}$. As seen in the upper right of Figure \ref{rotational_structure}, $P_{\rm RCB}$ will increase with the decrease in $f_{\rm p}$ . Especially, $P_{\rm RCB}$ would get higher with the spin speed. According to Equation (\ref{T_rcb}), the temperature at RCB would not be changed by the rotation, which only depends on the opacity we choose \citep{2015ApJ...811...41L}. Therefore, the temperature at RCB, even for a rotating planet, is approximate to $T_{\rm  RCB} \thicksim T_{\rm d}$. It is compatible with the result in the upper left of Figure \ref{rotational_structure}.

The locations of RCBs depend on the radiation energy densities of convective and radiative layers. Based on \cite{2015ApJ...803..111G}, the radiative energy density in the convective envelope follows $U=a_{\rm rad}T^4$. The optical depth satisfies $\tau = \int_{r}^{R}\kappa \rho dr$, where $a_{\rm rad}$ is the radiative constant and $R$ is the planet surface radius. Combined with Equation (\ref{hydrostatic_equilibrium}-\ref{thermal_gradient}),the relationship between the temperature and optical depth satisfies
\begin{equation}
\frac{{\rm d}T}{{\rm d}\tau} = \frac{g_{\rm r}T}{\kappa P}\nabla_{\rm ad} f_{\rm p}.
\label{dT_dtau}
\end{equation}
 Integrating this equation from the outermost radius of the envelope to the surface of the core, we find the temperature of core is
\begin{equation}
T_{\rm c} = \left[T_{\rm d}^{\beta} + \frac{\beta K^{n\left(a+1\right)}}{\kappa_{\rm 1} \left(k_{\rm B}/\mu m_{\rm u} K\right)^{\beta}} g_{\rm r} \nabla_{\rm ad} f_{\rm p} \tau_{\rm c}\right]^{1/\beta},
\label{Tc_tauc}
\end{equation}
where $\beta=\left(n+1\right)\left(a+1\right)+b$ with the polytropic index $n$,  while $K$ is polytropic constant and $\tau_{\rm c}$ is the optical depth of the core. The radiation energy density of core decreases with  $f_{\rm p}$, since $U_{\rm c} = a_{\rm rad}T_{\rm c}^4$ (see the solid lines of lower left of Figure \ref{rotational_structure}). Thus, rotation would decrease energy density of the convective layer.

The radiation energy density of the radiative envelope estimated by the diffusion approximation is
\begin{equation}
\frac{{\rm d}U}{{\rm d}\tau} = \frac{3}{c} \frac{L_{\rm int}}{4\pi R^2 }f_{\rm p}f_{\rm R},
\label{dU_dtau}
\end{equation}
where, $c$ and $L_{\rm int}$ denote  the speed of light and the internal luminosity of planet, respectively. Therefore, we can integrate Equation (\ref{dU_dtau}) from outer to the interior, and then find the radiation energy density of the radiative profile is
\begin{equation}
U_{\rm rad} = U_{\rm out} +\frac{3}{c} \frac{L_{\rm int}}{4\pi R^2 }f_{\rm p}f_{\rm R} \tau,
\label{U_rad}
\end{equation}
where, the radiation energy density of the planet's surface $U_{\rm out} $ is constant under the same boundary conditions. The coefficient $f_{\rm p}f_{\rm R}<1$ (see the lower left of Figure \ref{rotational_structure}).   There is an intersection, which is the radiative-convective transition point (RCB), between the radiative and convective profiles. Here, the optical depth at the RCB ($\tau_{\rm rad}$) can divide the envelope into two parts. The radiative gradient is smaller than adiabatic within the range of $\tau<\tau_{\rm rad}$, the envelope would be radiative. When  $\tau>\tau_{\rm rad}$, the adiabatic gradient becomes lower than the radiative, the envelope would be convective \citep{2015ApJ...803..111G}. Note that the radiative profile is linear in the optical depth, in which the slope corresponds to a tangent of convective profiles at RCB.
Besides, the tangent is decreased by rotation as the radiation energy density decreases. Therefore, the radiation energy density in the radiative layer decreases (see the dotted lines in the lower right corner of Figure \ref{rotational_structure}), resulting in the inward movement of the RCB (i.e., $\tau_{\rm rad}<\tau_{\rm rad,ro}$). Finally, we would obtain a higher mass and density at the RCB (i.e., $M_{\rm RCB}$, $\rho_{\rm RCB}$). We can predict that rotation could prolong the Kelvin-Helmholtz (i.e., KH) contraction timescale.

\begin{figure}
  \centering
  \includegraphics[width=8.5cm]{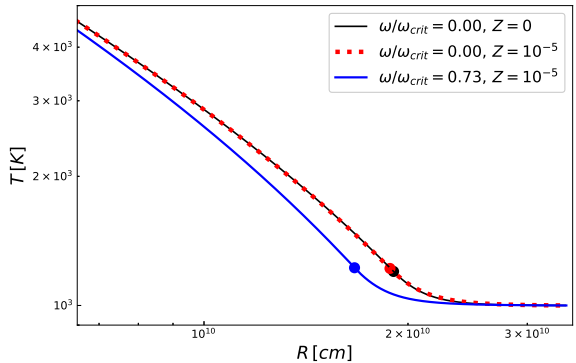}\vspace{-0.3cm}
\caption{The radial profiles of temperatures for three exemplary scenarios in which grain opacity enhancement has a different impact. The black line represents the fiducial model. The red dotted line represents the temperature that changes in the dusty atmosphere. The blue line corresponds to the combination of the dust grain and rotation. The dots denote the locations of RCBs. }
\label{dust}
\end{figure}

Dusty grains could provide a thick envelope to support the whole process of the evolution of the planet. These factors change the structures of rotating planets (see Figure \ref{dust}). In a dusty atmosphere, dust evaporation opens a radiative window that inhibits convection \citep{2014ApJ...797...95L}. Besides, the increase in opacity makes the temperature gradient of the radiative zone slightly steeper \citep{2020ApJ...896..135C} and forces the temperature at the RCB to increase. As a result, RCB will reach a greater depth. In Figure \ref{dust}, $T_{\rm RCB}$ increases in the thicker envelope of $Z = 10^{-5}$. As mentioned above, rotation can push the RCBs inward. Therefore, the interaction between spin and dust grain can compel the RCB to shift inward and increase $T_{\rm RCB}$ considerably.

\begin{figure}
 \centering
  \includegraphics[width=8.5cm]{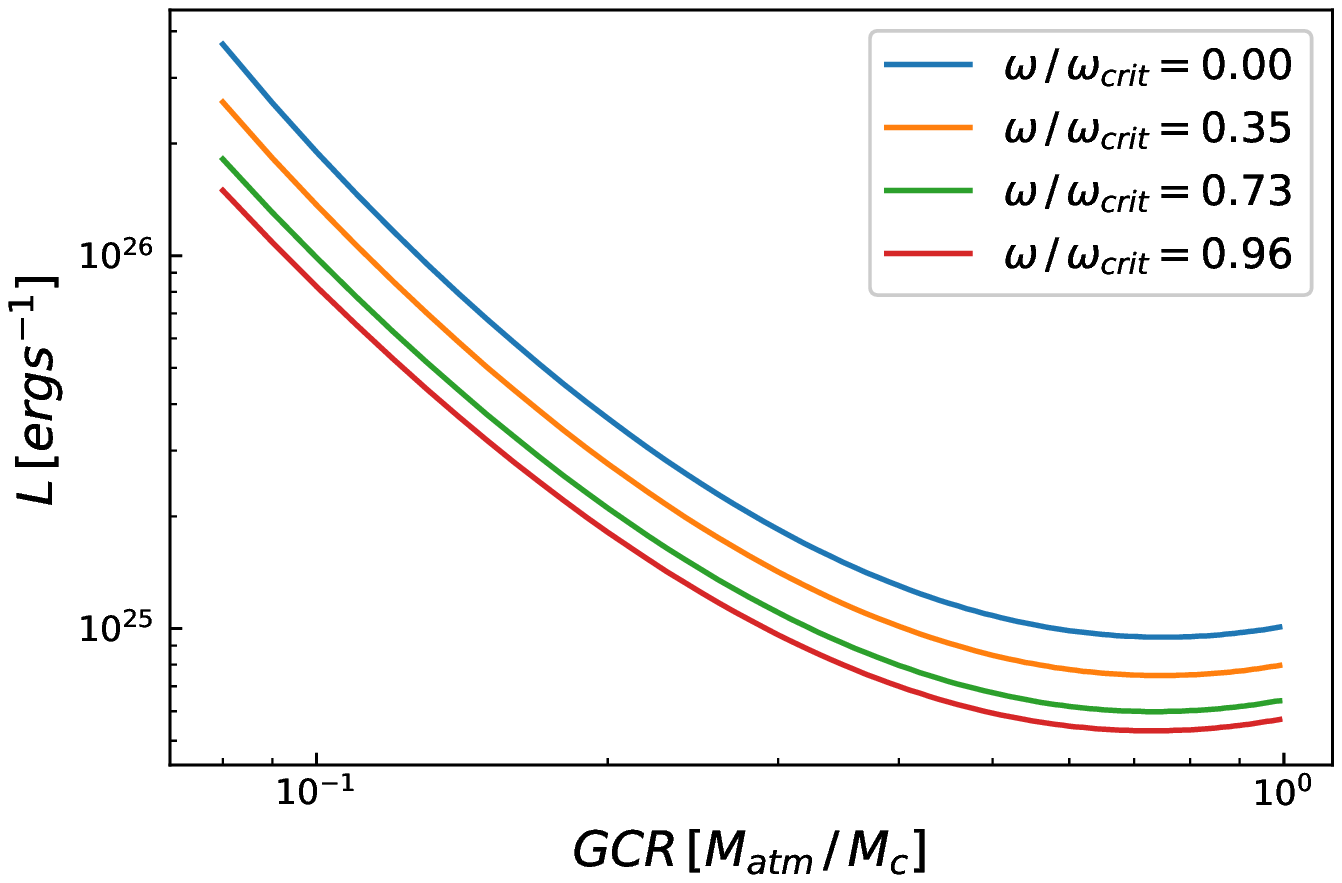}\vspace{-0.1cm}
  \includegraphics[width=8.5cm]{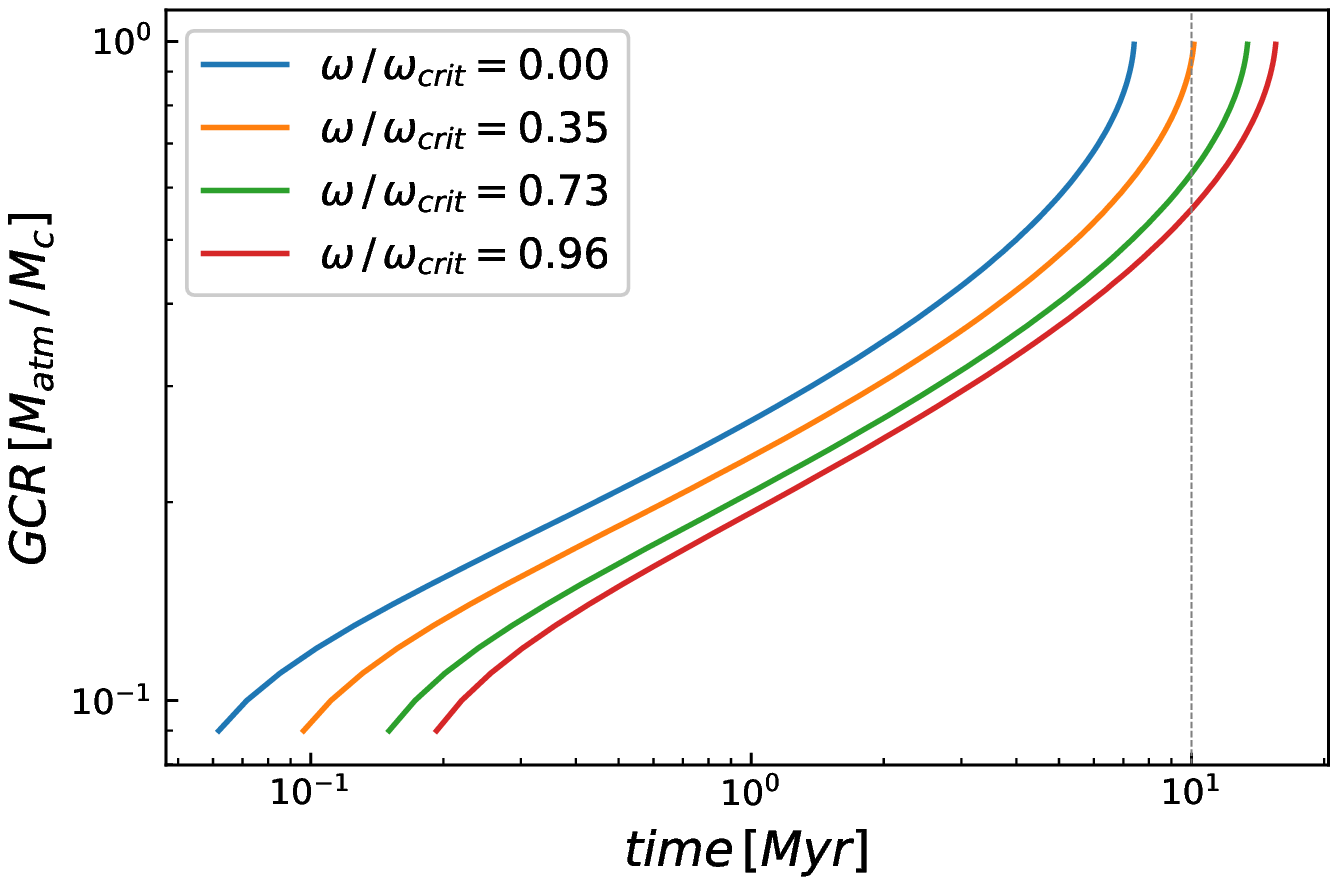}
  \caption{The upper panel: the luminosity as a function of gas-to-core mass ratio GCR for a  super-Earth/sub-Neptune under various spin speeds. The lower panel: the gas-to-core mass ratio GCR as a function of time for a  super-Earth/sub-Neptune. Blue, orange, green, and red lines correspond to the planet evolves with different angular velocities, i.e., $\omega/\omega_{\rm crit} =$0.00, 0.35, 0.73, and 0.96, respectively.}
\label{evolution_rotation}
\end{figure}

\begin{figure*}[!htp]
  \centering
  \includegraphics[width=8.5cm]{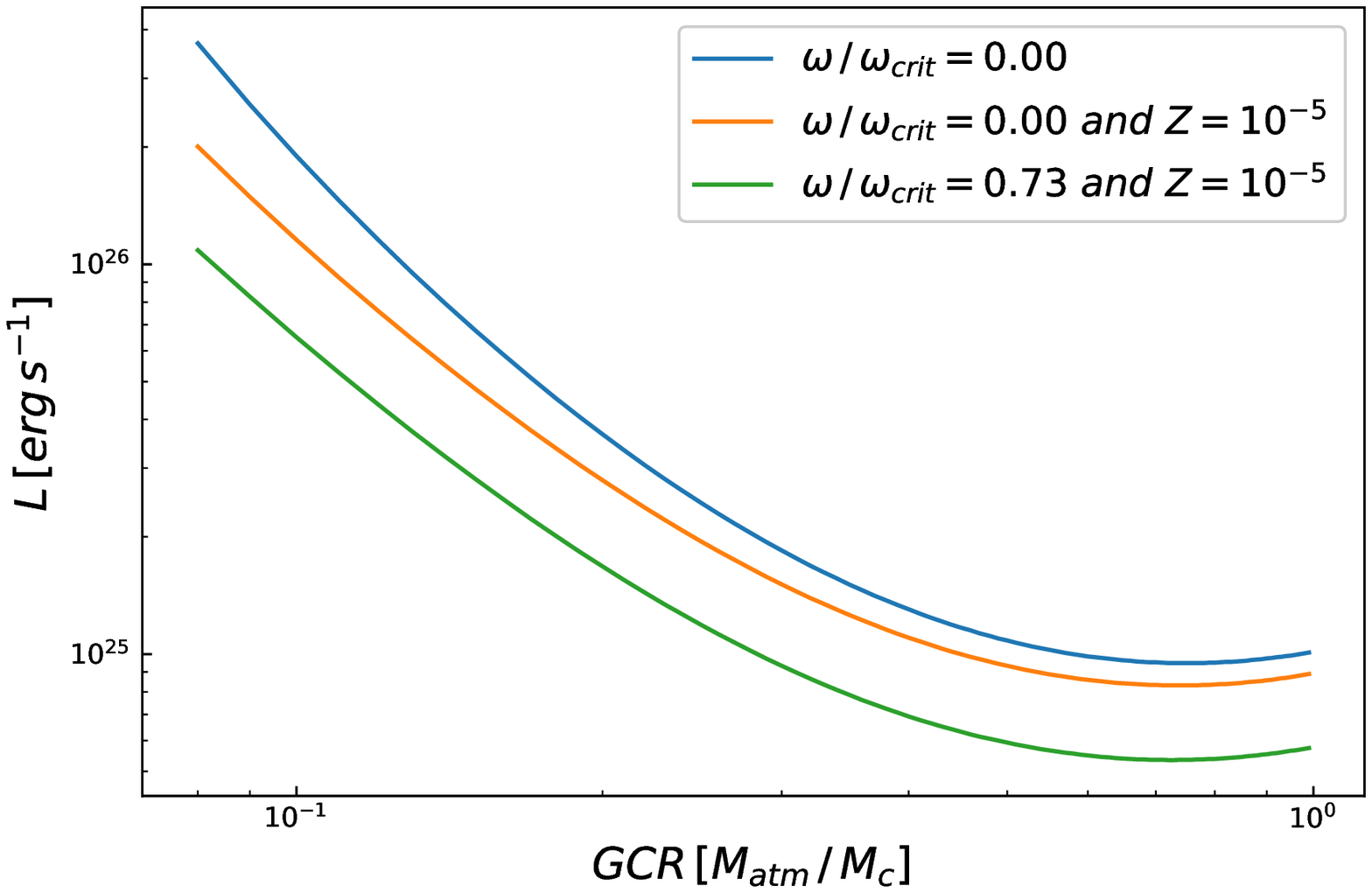}\vspace{-0.1cm}
  \includegraphics[width=8.5cm]{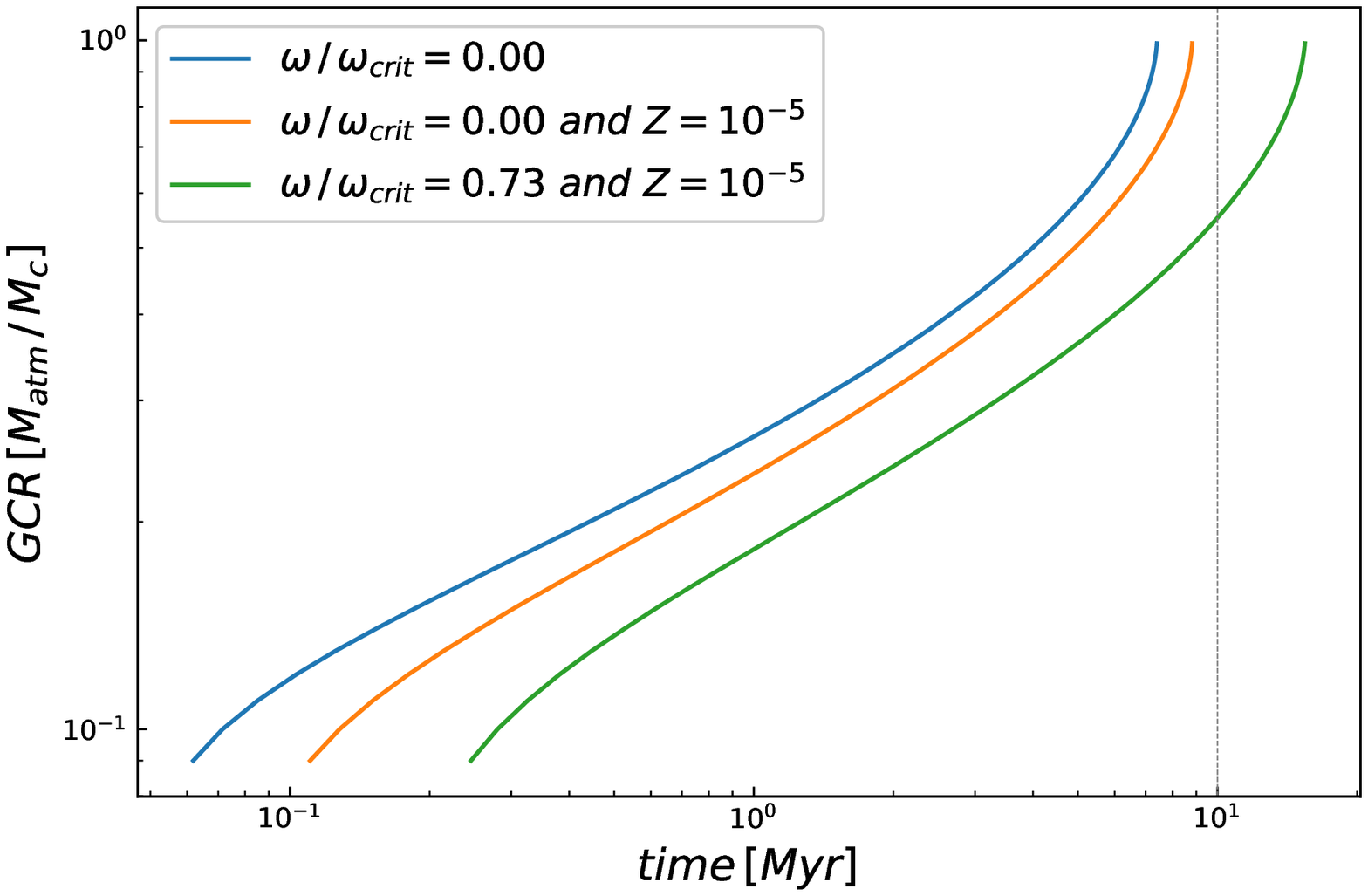}
\caption{The left panel: the relationship between the luminosity and  gas-to-core mass ratio GCR for a rotating planet without/with dust grain. The right panel: the gas-to-core mass ratio GCR as a function of time under different situations. The blue line corresponds to the fiducial model. The orange line shows the planet's mass grows in the dusty atmosphere, respectively. The green line corresponds to the combination of the dust grain and rotation.}
\label{evolution_dust}
\end{figure*}

\subsection{The Formation of  super-Earth/sub-Neptune}\label{Super-Earth}

To determine whether a planet grows to be a  super-Earth/sub-Neptune, we should understand the accretion history and calculate the evolutionary timescale. This section is divided into two parts. The formation of   super-Earths/sub-Neptunes  by rotation shows in Section \ref{rot_earth}. In Section \ref{rot_dust}, we consider the combination of dust grain and spin in forming  super-Earth/sub-Neptune.

\subsubsection{The Effects of Rotation}\label{rot_earth}

The accretion history is intimately related to the cooling luminosity. Therefore, it is necessary to investigate how the cooling luminosity varies with mass evolution. With the increase in the envelope mass, the steady-state envelope grows with the order of decreasing entropy \citep{2013prpl.conf2K014M}. This behavior complies with the cooling process of the envelope and allows the atmosphere to accrete more gas.

The cooling luminosity evolves with mass under different cases, shown in the upper panel of Figure \ref{evolution_rotation}. In the early phase, the luminosity decreases with increasing mass. The self-gravity effect does not play an essential role in this early phase, and luminosity diminishes with a thicker radiative outer layer and a massive envelope \citep{2017ApJ...850..198Y}. When the self-gravity effect is ignored, the luminosity at the RCBs is 
\begin{equation}
\begin{aligned}
L_{\rm RCB}= &\frac{64 \pi \sigma G M_{\rm p} T_{\rm RCB}^4}{3\kappa P_{\rm RCB} f_{\rm R}}
 \nabla_{\rm ad} \\
 \propto & \frac{ T_{\rm RCB}^4}{\kappa P_{\rm RCB} f_{\rm R}}.
\end{aligned}
\label{luminosity_rcb}
\end{equation}
Note that $P = k_{\rm B} \rho T/\mu m_{\rm u}= K\rho^{1+1/n}$ at RCB and $\kappa = \kappa_{\rm 0}\left(P/P_{\rm 0}\right)^a \left(T/T_{\rm 0}\right)^b $.  Therefore, Equation (\ref{luminosity_rcb}) can
switch to
\begin{equation}
L_{\rm RCB} \propto  T_{\rm RCB}^{4-b-\left(n+1 \right)\left(a+1 \right)}/ f_{\rm R}.
\label{L_new_rcb}
\end{equation}
As $g_{\rm eff}$ is remarkably less than $g_{\rm r}$, $f_{\rm R}>1$ according to Equation (\ref{f_r}). Since $T_{\rm RCB} \sim T_{\rm d}$, the luminosity would decrease with the increase in the coefficient $f_{\rm R}$. As seen in the upper panel of Figure \ref{evolution_rotation}, The planet grows with the decreases in luminosity.  The rotation would reduce luminosity, which decreases with the increase in angular velocity. 

A gas giant can form in a gas-rich disk when the planet enters the runaway accretion stage, meaning it grows within the disk lifetime. However, if the evolutionary timescale exceeds the disk lifetime, the disk gas is depleted. The envelope can not accrete sufficient gas to be a gas giant.  Subsequently, the planets except for gas giants will enter the post-formation  evolution, the primal envelope would experience a mass loss. If the mass loss is slight, the planet would turn into a sub-Neptune. If the envelope significantly loses its mass, the super-Earth would be formed.  Therefore, two timescales, the evolutionary timescale ($t$) and the disk lifetime ($t_{\rm disk}$), can predict whether a plant becomes to be a super-Earth/sub-Neptune. As $t>t_{\rm disk}$, a super-Earth/sub-Neptune may form in future. Note that the disk time is between 3-10 Myr. To specify, we take disk lifetime as $t_{\rm disk}=10$ Myr in this work.


The relationship between the evolutionary timescale and GCR shows in the lower panel of Figure \ref{evolution_rotation}.
The planet’s atmosphere accretes materials from the proto-planetary disk, we can estimate the evolutionary timescale by \citep{2013Stellar,2015ApJ...811...41L}:
\begin{equation}
t \thicksim  \frac{|E| }{L}\thicksim \frac{GM^2}{RL} \propto T_{\rm RCB}^{ N}f_{\rm R}/R_{\rm RCB},
\label{new_time}
\end{equation}
where, ${ N} ={\left(n+1\right)  \left(a+1\right)+b-4}>0$.  As mentioned above, the rotation would push RCB inward and then decrease the luminosity. As a result, the KH timescale will be prolonged, according to Equation (\ref{new_time}). Besides, the fast angular velocity will enlarge this signature. As we can see, the evolutionary timescale would get higher than $10$ Myr. A  super-Earths/sub-Neptunes would form in this state.

\subsubsection{The Combination of Dusty Grain and Rotation }\label{rot_dust}

 Following Lee \& Chiang 2015, the temperature at the RCB in a dust-free atmosphere satisfies the relation $T_{\rm RCB}\sim T_{\rm d}$, and the runaway timescale is directly proportional to the opacity $t_{\rm run} \propto \kappa$. However, $T_{\rm RCB}$ in the dusty atmosphere is around the hydrogen molecular dissociation front ($\sim 2500\,K$), as the thermal evaporation of dust within the envelope pushes the RCB inward. We also have verified the dusty grain would change the parameters at RCBs. Thus, when we take both rotation and the dusty grain into account, the evolutionary timescales will show a difference. In this work, the inputting dust grain will increase the planet's opacity due to $\kappa_{\rm total}= \kappa_{\rm gas}+\kappa_{\rm gr}$ \citep{2014Ormel}. 
Following Equation (\ref{luminosity_rcb}), the cooling luminosity switches to
\begin{equation}
L_{\rm RCB} \propto \left[\kappa_{\rm 1} \Gamma^{a+1} T_{\rm RCB}^{N}  +\kappa_{\rm gr} \Gamma T_{\rm RCB}^{\left(n+1\right)-4} \right]^{-1}f_{\rm R}^{-1},
\label{l_rcb_dust}
\end{equation}
where, $\Gamma=K \left(k_{\rm B}/\mu m_{\rm u} K\right)^{n+1}$. Especially, $\kappa_{\rm gr} = 0$ represents the planet is formed without the dusty grain. The evolutionary time consists with $t \propto \left(R_{\rm RCB} L_{\rm RCB}\right)^{-1}$. As mentioned in Section \ref{three_layer}, the temperature at RCB, accompanying the atmosphere thickening, would increase (i.e., $T_{\rm RCB} > T_{\rm d}$).   Besides,  the RCB compared to the fiducial model would penetrate deeper, and it will be deepened by rotation. Consequently, they can reduce the cooling luminosity (see the upper panel of Figure \ref{evolution_dust}) and enhance the KH time with the gravitational coefficient $f_{\rm R}$.  Thus, dusty grain can help   super-Earth/sub-Neptune  form with the lower speed of rotation.

\section{conclusion and discuss}\label{sec:conclusion}

The massive core may accrete extended atmospheres and grow to be the gas giant and gas-rich super-Earth \citep{2019MNRAS.488.2365B}.  Due to the conservation of vortensity,  the amount of rotational support is acquired for the envelope as the core mass increases more than the thermal mass \citep{2019MNRAS.487.2319B}.  
However, the measurement for planetary rotation is scarce except for Jupiter and Saturn. 
\cite{2020ApJ...905...37B} measured the rotating line of eight planetary-mass objects and found the velocity is inversely proportional to the radius as the planets within the conservation of the angular momentum cool and contract. There are many uncertainties, such as gravity and shape, in the process of locating RCBs and determining the composition of rotating planets. The rotation affects the centrifugal force and deformation in the development of planets, thus the evolution of planets is altered.

In this paper, we mainly discuss how planetary rotation affects the formation of  super-Earths/sub-Neptunes. RCB penetrates deeper by spin since the radiation energy densities in the radiative and convective envelope decrease. The temperature at RCB is consistent with $T_{\rm RCB}\sim T_{\rm d}$, and the pressure at RCB increases. Consequently, the evolutionary timescale grows with a reduction in luminosity and a super-Earths/sub-Neptunes forms when the evolutionary timescale $t >10$ Myr. In a thicker envelope (with the dusty grain), the rotation would amplify these differences. Therefore, by assuming that dusty grain in a rotating planet can extend the evolutionary timescale, we can also predict that a super-Earths/sub-Neptunes may eventually form. In this work, a super-Earths/sub-Neptunes can generate in the inner disk region by rapid rotation. It may not occur in the outer disk since the evolutionary time and the orbital speed decrease with the orbital radius.

Rotation can well explain the interior structure of a gas giant with the measurements data. There is a lack of understanding of how a  super-Earths/sub-Neptunes forms by rotation. However, this work also suggests a super-Earths/sub-Neptunes could be formed through spin within a workable mass range. Rotation pushes the RCB inward and then enlarges the radiative region. Besides, rotation in the envelope may cause fluid turbulence and induce vorticity \citep{2019MNRAS.487.2319B,2021Icar..36114394B}. Correspondingly, the changes of gravitational energy would relate to the induced potential vorticity flux.
Thus, spin may noticeably enlarge the radiative zone of a planet. 



A limit of this work is that the formation of planets under the condition of no special assumption about the angular velocity. When the angular speed is not specified, it introduces the distributions of the spin speed $f_{\rm \omega}$ and potential $f_{\rm \psi}$ in the planetary structure (see the appendix of \citealt{2002A&A...394..965Z}), resulting in the complex calculation. As a compromise, we tried to parametrize the value of angular velocity as a free parameter to control the angular velocity $\omega<\omega_{\rm crit}$ within a feasible range. When $\omega=\omega_{\rm crit}$, the gravitational energy equals centrifugal energy, and then the planet breaks up.

Generally, rotation can be treated as a two-dimensional (i.e., 2D) model for oblate-ellipsoids \citep{1986ApJS...61..479H}. Based on the two-dimensional simulations, we can research the effects of several instabilities on the formation of gas giants or super-Earths, such as west winds, jet streams, or the various effects of the Coriolis force driven by rotation or differential rotation \citep{2009PhT....62i..52M}. For instance, on the fast rotating surface of Jupiter, turbulent waves at the interface of differential rotating zones created the Red Spot, which is a perpetual hurricane generated by planetary rotation \citep{2009PhT....62i..52M}. The loss of angular moment, driven by rotation, dominates meridian circulation and turbulence \citep{1992A&A...265..115Z}. Therefore, taking other instabilities into an accretion model may reveal some interesting features.

\section{acknowledgments}
We thank the referee for helpful comments. This work has been supported by National Key R \& D Program of China (No. 2020YFC2201200) and the science research grants from the China Manned Space Project (No. CMS-CSST-2021-B09 \& CMS-CSST-2021-A10).  C.Y. has been supported by the National Natural Science Foundation of China (grants 11373064, 11521303, 11733010, and 11873103), Yun-nan National Science Foundation (grant 2014HB048), and Yunnan Province (2017HC018).

\bibliography{sample631}{}
\bibliographystyle{aasjournal}

\end{document}